\documentclass[aip,pop,reprint]{revtex4-1}
\usepackage{graphicx}
\usepackage{amsmath}
\usepackage{subfigure}
\begin{document}

\title{Many-body collision contributions to electron momentum damping rates in a plasma influenced by electron strong coupling}
\author{Puchang Jiang}
\author{John Guthrie}
\author{Jacob L. Roberts}%
\affiliation{%
 Physics Department, Colorado State University 
}%

\begin{abstract}
Experimental studies of electron-ion collision rates in an ultracold neutral plasma (UNP) can be conducted through measuring the rate of electron plasma oscillation damping. For sufficiently cold and dense conditions where strong coupling influences are important, the measured damping rate was faster by 37\% than theoretical expectations [W. Chen, C. Witte, and J. Roberts, Phys. Rev. E \textbf{96}, 013203 (2017)]. We have conducted a series of numerical simulations to isolate the primary source of this difference. By analyzing the distribution of electron velocity changes due to collisions in a molecular dynamics simulation, examining the trajectory of electrons with high deflection angle in such simulations, and examining the oscillation damping rate while varying the ratio of two-body to three-body electron-ion collision rates, we have found that the difference is consistent with the effect due to many-body collisions leading to bound electrons. This has implications for other electron-ion collision related transport properties in addition to electron oscillation damping.
%plays an important role of understanding the fundamental plasma physics.Experiment and variety of computation method such as Molecular dynamics(MD) simulation agreed on the measurement of electron oscillation damping rate thus indicate the same electron ion collision rate. But calculation results from simulation which treats the ions as a background potential and only simulate electron dynamics under such potential disagreed with the MD simulation and experiment data. In this paper we discuss a possible explanation to this discrepancy and develop a new model to compensate the gap.

\end{abstract}

\maketitle

%\section{Introduction}

In strongly coupled plasmas, the average nearest-neighbor Coulomb potential energy for one or more types of particles is comparable to or exceeds the average kinetic energy of that type of particle. As a result, spatial correlations between plasma particles become significant and many common assumptions used in plasma theory break down.\cite{ichimaru1982strongly,murillo2015ultracold,chen2017observation,bannasch2012velocity,strickler2016experimental,murillo2015ultracold,haenel2017arrested,chen2004electron} In contrast to weakly-coupled plasmas where collisions can be treated as long-range binary scattering events, in strongly-coupled plasmas, many-body effects become increasingly significant. These many-body effects can be treated theoretically through, for instance, using simulations to create phenomenological extensions of binary collision theory into the strongly-coupled regime,\cite{stanton2016ionic} or through creating an effective potential that incorporates many-body effects into inter-particle effective potentials.\cite{baalrud2014extending}  

However, in strongly-coupled plasmas with electrons and ions (as opposed to one-component plasmas), three- and more-body collisions can result in electrons becoming bound to one or more ions in the plasma, forming a Rydberg atom or an electron bound to a small number of ions. \cite{fletcher2007using,bannasch2011rydberg,mansbach1969monte, robicheaux2003simulated}
%For instance, two electrons and an ion can collide to form an atom (and leave a remaining free electron).
In this article, we show that such many-body collisions 
%are not only relevant from an atom-formation rate perspective, but also 
provide a substantial contribution to the electron average momentum damping rate in plasmas where electron strong coupling is significant. For such strongly coupled plasmas, these many-body-to-bound-state (MBTBS) collisions would be expected to be relevant for other collision-related properties as well, such as stopping power,\cite{zylstra2015measurement,frenje2015measurements} thermalization rates,\cite{dimonte2008molecular} and other transport properties. \cite{hinton1976theory}

%The reason that this is the case is that bound electrons often do not remain bound as the atom can be re-ionized after some time, and so both the fact that these bound states exist and the net electron motion in these bound states are important for the net motion of the electrons.  These three-body collisions can thus be expected to have significance for transport properties, stopping power, and thermalization in plasmas in a way that is not amenable to a binary collision treatment in strongly coupled plasmas.

This investigation of MBTBS collision contributions to the average electron momentum damping rate was motivated by studies of electron oscillation damping in an ultracold neutral plasma (UNP).\cite{chen2017observation} In that work, electron center-of-mass oscillations were induced by imparting an impulse acceleration in one direction to the electrons' velocities. The electrons' center-of-mass then oscillated with respect to the ions until that oscillation damped out due to electron-ion collisions. At the highest electron strong coupling condition measured, 
%For conditions where the electrons’ strong coupling parameter $\Gamma=\frac{(\frac{4\pi n}{3})^{\frac{1}{3}}e^2}{4\pi \epsilon_0 k_B T_e}=0.35$,
there was a gap between the predicted damping rate obtained from multiple theories and the observed damping rate. This gap was also present between a theoretically predicted damping rate and the rate obtained in a molecular dynamics (MD) simulation that agreed with the experimentally measured rate.\cite{chen2017observation} The ratio of the simulated to predicted damping rate was 1.37 at the coldest electron temperatures that were considered. To identify the source of the gap, we conducted more simulation and theoretical numerical studies to identify the source of the observed discrepancy.  The result of those studies is described in this work.

UNPs are formed through photoionizing cold atoms or molecules from a laser-cooled gas\cite{killian1999creation} or supersonic beam\cite{morrison2008evolution} in a finite-spatial-extent volume. Once the atoms or molecules are photoionized, some electrons escape from the UNP formation region as the average space charge is initially neutral. The more massive ions remain, however, and so a positive space charge develops that is eventually sufficient to confine the remaining electrons, forming the UNP.\cite{killian1999creation} Both the ion and electron components quickly come to thermal equilibrium individually, but not with each other. The UNP thus consists of a finite-spatial-extent plasma of electrons and singly-charged ions where the electrons and ions are individually in thermal equilibrium. The electron temperature can be controlled via the photoionization laser wavelength.\cite{killian1999creation} Eventually, the UNP expands since the ions are not confined,\cite{kulin2000plasma} but the timescale of this expansion is long enough that it is insignificant for the work described in this article.

Three-body recombination (TBR) to Rydberg atoms plays an important role in UNPs with sufficiently high electron strong coupling parameter $\Gamma$.\cite{bannasch2011rydberg, robicheaux2002simulation} The electron strong coupling parameter is defined as the ratio of the nearest-neighbor Coulomb potential energy to thermal energy such that $\Gamma=\frac{e^2}{4\pi \epsilon_0 a_{\mathrm{WS}}}/k_\mathrm{B} T_e$
where $e$ is the fundamental electron charge, $\epsilon_0$ is the dielectric constant of the vacuum, $k_\mathrm{B}$ is Boltzmann's constant, $T_e$ is the electron temperature, and $a_{\mathrm{WS}}$ is the Wigner-Seitz radius equal to $(\frac{3}{4\pi n})^{1/3}$ where $n$ is the electron density. TBR occurs when two electrons collide near an ion and one of the electrons becomes bound, forming an atom that is overwhelmingly likely to be in a highly excited state --- a Rydberg atom. Not only are a free ion and free electron lost from the plasma, but the binding energy of the atom is imparted to the non-bound electron involved in the collision and that ultimately results in heating of the electron component of the plasma.\cite{robicheaux2002simulation,bannasch2011rydberg} In the weakly-coupled limit, three-body rate scales strongly with electron temperature as $T_e^{-9/2}$. This Rydberg atom formation heating mechanism is predicted to limit the highest achievable values of electron $\Gamma$ in UNPs.\cite{robicheaux2002simulation}

Rydberg atoms formed in this way, however, do not remain at a fixed binding energy. Electron-Rydberg atom collisions can change the Rydberg atom energy. These collisions can also ionize the Rydberg atom.\cite{bannasch2011rydberg} Electrons thus move in and out of being bound. Moreover, it is possible for electrons to be bound not only to a single ion, but to multiple ions. We found that weakly-bound Rydberg electrons and electrons bound to multiple ions (which are necessarily weakly bound) made the largest many-body contributions to the collision and damping effects described here.

If electrons are moving with an average velocity relative to ions, then electron-ion collisions will result in a damping of that velocity.
%This occurs because the electron velocities are deflected as a result of the electron-ion collision, reducing the electron center-of-mass speed.
For a bound state electron, the electron's velocity direction changes rapidly as compared to the rate of change of a free electron colliding with other electrons and with ions. An electron that becomes bound and then re-ionizes after a sufficiently short period of time will thus undergo a much larger velocity deflection on average than if it had remained as a free electron. These electrons have a disproportionate impact on the damping of any average electron velocity with respect to the ions. Even if they are only a small fraction of the number of electrons in the plasma, because of comparatively large velocity deflections they can play a significant role in the damping rate.

While the above considerations indicate that MBTBS collisions will have some effect on the average electron momentum damping rate, they do not quantify its contribution. In the rest of this article, numerical simulations are described that demonstrate that for the UNP plasma conditions of Ref. \onlinecite{chen2017observation}, the MBTBS collisions are primarily responsible for the gap between predicted and observed/simulated electron oscillation damping rates.

%\section{\label{sec:level1}introduction}
%\section{\label{sec:level1}Three-body recombination background}

%There are several length scales describing UCPs. Debye screening length $\lambda$  is defined by $\lambda=\sqrt{\epsilon_0 k_b T/e^2 n}$ where $e$ is the electron charge, $\epsilon_0$ is the dielectric constant of free space, $n$ is plasma density, $k_B$ is Boltzmann's constant. Impact parameter $b$ is defined by $b=e^2/4\pi \epsilon_0 k_b T$. Inter-particle spacing $a$ is defined by 
%\section{\label{sec:level1}The utlracold plasma system and electron oscillations in that system}
\section{Numerical Simulation of Electron Center-of-mass Oscillation Damping in a UNP}

%The formation of ultracold plasma evolves with standard laser cooling and photoionization technique.Rubidium atoms are laser cooled to about 100mK in ultra high vacuum chamber. Then lasers with certain frequencies are applied to the atoms to excite lower energy state electrons to ionized state. The ionized electron temperature can be tuned by the laser frequency. After ionization, electron temperature will change under the influence of different cooling and heating mechanism such as evaporative cooling, adiabatic cooling, disorder-induced heating, three body recombination heating, continuum lowering heating. Without any electric or magnetic field, electrons will soon reach to equilibrium on the order of microseconds scale determined by the plasma density. For our ultracold plasma in both experiment and computer simulation, the initial electron temperature is on the order of 1K, plasma size is about 1000 micron with 16000 electrons and 30000 ions. Under such condition, UCP lifetime is on the order of 1ms and it takes several hundreds of microsecond for the electrons to equilibrate.

Through simulations we calculate the UNP electron center-of-mass oscillation damping rate and other electron properties using two separate techniques to investigate relative contributions from binary and MBTBS electron-ion collisions. In the first, we perform what we term a full molecular dynamics (MD) simulation. The full MD simulation has been shown to agree with experimental measurements and should contain all the relevant physics. We call the second technique the Monte Carlo (MC) simulation. The MC simulation eliminates explicit electron-ion collisions by not representing the ions as point particles. The electron-ion collisions are instead modeled via a random binary collision operator. Given a theory for electron-ion collisions, the collision operator can be constructed.  This allows for a translation of any electron-ion collision theory to any UNP quantity of interest, calculated via the MC simulation. These two techniques will be described in more detail in turn below.

In the full MD simulation, the electrons and ions are treated as point particles that interact pairwise via a modified Coulomb potential defined to be $U=\frac{q_1q_2}{4\pi \epsilon_0}\frac{1}{\sqrt{r^2+\alpha^2}}$, where $q_1$ and $q_2$ are the charges of the interacting particles, $r$ is the interparticle distance and $\alpha$ is a constant. The motivation for adding the $\alpha$ term will be discussed below. In each timestep, the net force on each particle from all other particles is calculated to generate an acceleration. The motion of the particles is integrated by using a Leapfrog method.  
%At every timestep, the position and velocity is known for every electron and ion and these positions and velocities are periodically recorded.

The $O(N^2)$ scaling of the force calculation introduces significant computational demands. While algorithms are available that reduce the computational scaling with particle number,\cite{dharuman2017generalized,kudin1998fast} we address this challenge by using a hardware solution. The force computations are easily made parallel and are performed on a GPU using a software architecture based on the OpenCL framework.

The addition of the constant $\alpha$ is necessary to prevent numerical problems that arise from deeply-bound Rydberg atoms that can form in the UNP. The velocity of the electrons in such Rydberg atoms is high, and so for practical timesteps numerical errors accumulate and energy is no longer conserved. ``Softening" the Coulomb potential with the $\alpha$ term mitigates this problem.\cite{tiwari2018reduction} Given the importance of Rydberg atoms in this investigation, the possible distortion of Rydberg populations through the addition of the $\alpha$ term is a concern. We chose $\alpha$ to be equal to just under 2\% of $a_{\mathrm{WS}}$. We made sure that we could vary $\alpha$ around this value in a range that spanned about a factor of two without affecting any of the calculated quantities of interest, such as the oscillation damping rate, in any detectable way. While the most deeply bound Rydberg population number will not be correct, those Rydbergs were found not to have a significant contribution to the MBTBS-related effects and so the use of $\alpha$ did not distort the physics of interest.

In the second type of simulation, the MC simulation, the electrons are still treated as point particles, acting through the same modified Coulomb potential for consistency (with $\alpha$ again having no measurable impact). The ions, however, are not represented as point particles. Instead, a continuous charge distribution equal to the average ion density as a function of position is used. This provides the confining potential for the electrons. Electron-ion collisions are modeled through a random collision operator.

In each timestep, a collision probability is calculated for each electron based on the ion density at the electron's position, the electron's velocity, and the total cross section associated with the collision operator. A random number is generated and compared to the collision probability to determine if a collision has occurred. If one does occur, then the electrons' velocity is deflected randomly according to a distribution function that is generated from the underlying theory of electron-ion collisions that is being used.

A few simplifying assumptions were made in both the full MD and MC simulations. A uniform ion density distribution was used so that the ion density was constant. Aside from convenience, the uniform ion density eliminated collisionless oscillation damping that can occur in non-uniform-density UNPs.\cite{chen2016damping} The electron number was kept smaller than the ion number to generate confinement. In the full MD simulation, the ion mass was increased to be effectively infinite, as this removed any possible expansion or ion-ion correlation effects.  Simulations with the actual $^{85}$Rb mass used in the experiment\cite{chen2017observation} showed these effects to have effects smaller than a few percent, at most, justifying the infinite mass approximation that was made.  

Before any properties were extracted from the simulations, they were initialized. For the full MD case, electrons and ions were placed in random locations given the uniform density. Typically, 16,000 electrons and 30,000 ions were used with an average density of $1\times10^{13}\,\mathrm{m}^{-3}$. The simulation was then allowed to run for $0.5\,\mathrm{\mu s}$ in order to let the electrons come to equilibrium. The electron temperature is calculated by assuming $\frac{3}{2}N_e k_\mathrm{B} T_e=\sum_{i=1}^{N_e}\frac{1}{2}mv_i^2$ where $m$ is the electron mass, $N_e$ is the electron number, and $v_i$ is the speed of the $i^\mathrm{th}$ electron. The electron temperature would increase during the equilibration period owing to disorder-induced-heating\cite{kuzmin2002numerical,murillo2001using,lyon2011influence,chen2004electron} and TBR\cite{robicheaux2002simulation,killian2001formation,bannasch2011rydberg}. To tune the temperature to our desired value, midway through the equilibration we would rescale all electron velocities to produce the desired total kinetic energy and then we would let the system equilibrate the rest of the time. The MC initialization proceeded in the same way, only without the need to include point charge ions.

\begin{figure}
    \centering
    \includegraphics[width=0.48\textwidth]{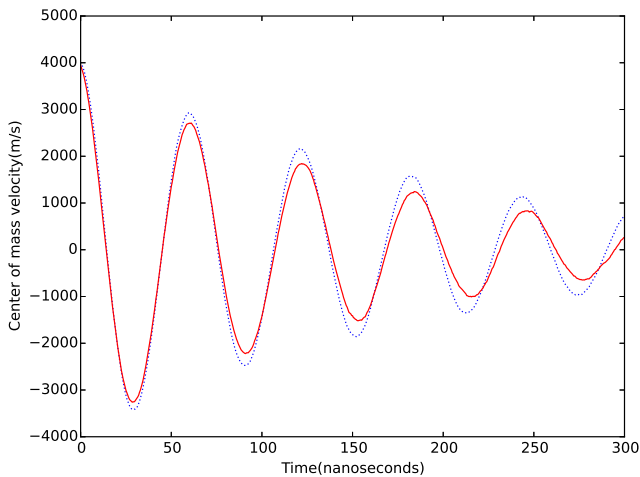}
    \caption{Center-of-mass oscillation damping numerical results. The electrons' center-of-mass velocity as a function of time is plotted, showing the oscillation decay. The red solid line is obtained from the full MD simulation while the blue dotted line is from the MC simulation. For this simulation, the electron temperature was $T_e=1.65\,\mathrm{K}$ with the other relevant conditions described in the main text.}
    \label{fig:osc_data_sample}
\end{figure}

Once equilibrium had been established, we would generally perform one of two types of simulation runs. In the first type, we would simulate the oscillation damping. To do so, we would instantaneously increase all of the electrons' velocities by $4.00\,\mathrm{km/s}$ in the z direction. The electron center-of-mass would then oscillate. To measure the oscillation damping time, we would fit a damped sine curve to the oscillation data to extract the damping time constant. See Fig. \ref{fig:osc_data_sample} for an example of the results obtained from this type of simulation run. Results could be obtained from either the full MD or MC simulations.

In the second type of simulation run, we would not alter the electron velocities after equilibration, but would just record them at evenly-spaced time intervals to measure their individual velocity deflections. The motivation for doing this is that the practical effect of electron-ion collisions is to increase the rate that electrons deflect from their initial velocity direction. To characterize the amount of deflection over a chosen time period, the deflection angle $\Delta \theta$ for each individual electron was calculated. The distribution of these deflection angles could be compiled and compared. If the collision operator in the MC simulation did a perfect job of modeling electron-ion collisions, then the distribution of deflection angles between the MC and full MD simulations would be the same and the oscillation damping rate would be identical, too. Any differences in deflection angle distributions would indicate that the full MD includes physics that is not being captured by the MC simulation.

We measured deflection angle distributions over a wide range of time periods, but concentrated our studies on the deflections over 5, 10, and 20 $\mathrm{ns}$. The 5 $\mathrm{ns}$ time period is approximately equal to the time for an electron to move $a_{\mathrm{WS}}$. We note that electron-electron collisions will cause deflections, too, and so even in the absence of electron-ion collisions there will be a range of electron deflection angles. In fact, electron-electron collisions not only deflect but also alter the speed of electrons.

\begin{figure}
    \centering
    \includegraphics[width=0.48\textwidth]{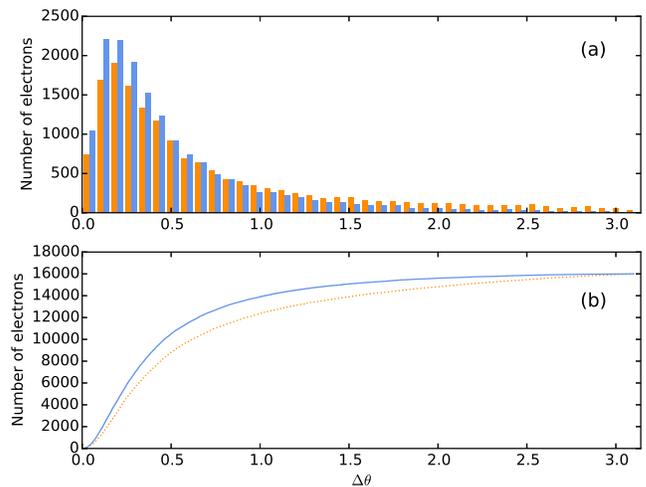}
    \caption{Comparison of the distribution of electron velocity deflection angles over a time period of 5 $\mathrm{ns}$. For reference, $\omega_p^{-1}=\sqrt{\epsilon_0 m/e^2 n}=5.60\,\mathrm{ns}$. In Fig. \ref{fig:histrogram}(a), The orange bars correspond to full MD simulation results, while the blue bars are obtained from MC simulations. In Fig. \ref{fig:histrogram}(b) the orange dotted line represents MD simulation while blue solid line represents MC simulation results. For these simulations, $T_e=1.65\,\mathrm{K}$. Similar trends were observed for deflections computed over longer time periods than 5.0 $\mathrm{ns}$. The MC collision operator used is described in the main text.}
    \label{fig:histrogram}
\end{figure}

A natural way to plot distributions of deflection angles $\Delta \theta$ is through using a histogram, as shown in Fig. \ref{fig:histrogram}(a). However, we found that an alternative way of displaying the same results was useful. Such a plot is shown in Fig. \ref{fig:histrogram}(b). In this plot, the x-axis runs through all possible deflection angles. The y-axis plots the number of electrons with a deflection angle equal to or less than the associated deflection angle. In Fig. \ref{fig:histrogram}(b), the full MD curve has a smaller number of electrons through moderate deflection angles, indicating that there are more high-angle deflections in the full MD results than the MC results. This fact is also apparent in Fig. \ref{fig:histrogram}(a).

\begin{figure}
    \centering
    \includegraphics[width=0.48\textwidth]{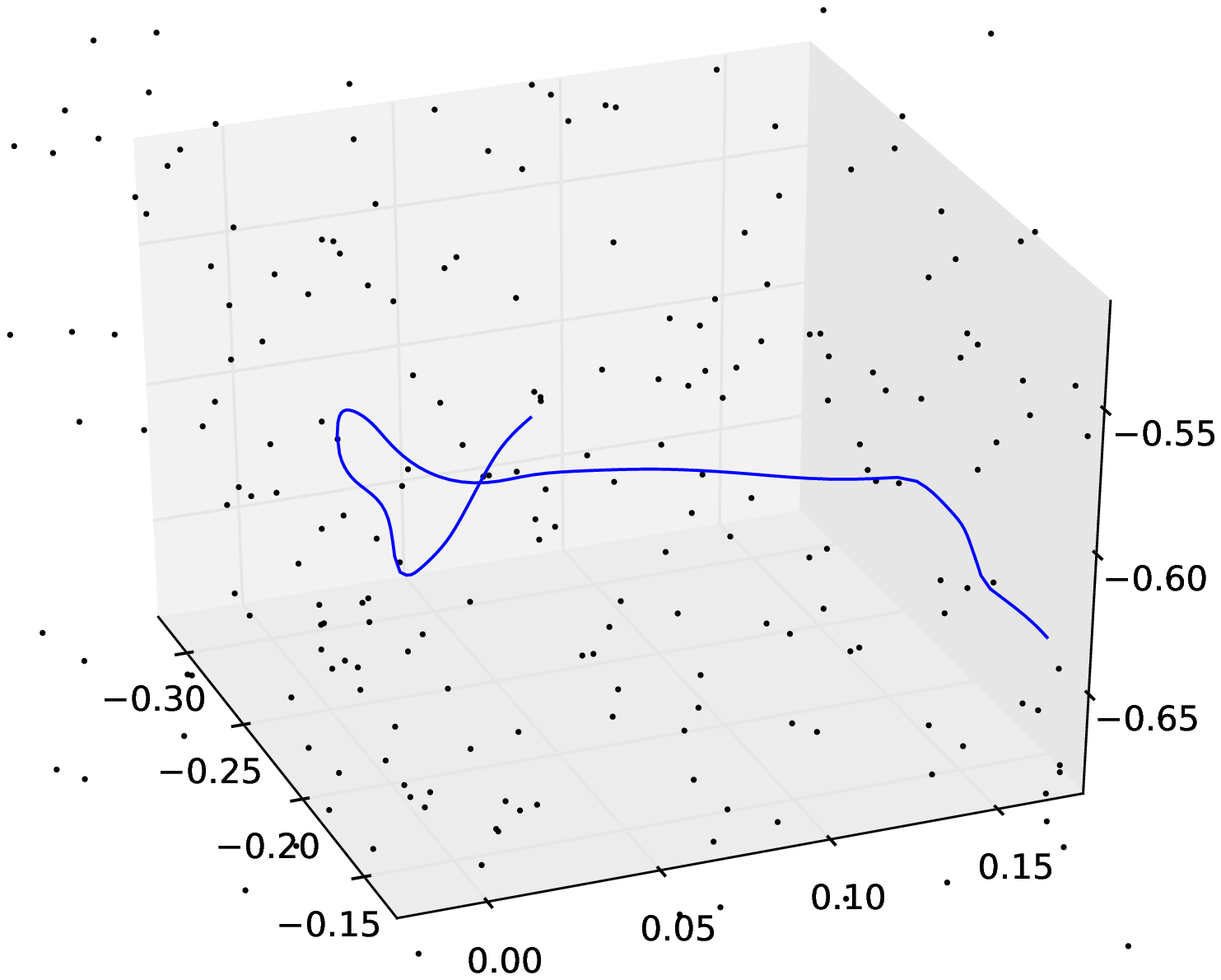}
    \caption{Sample electron trajectory. This 3D plot shows an electron trajectory over a period of 60.0 $\mathrm{ns}$. This electron trajectory was chosen to illustrate an electron that is initially free, but then becomes bound to a cluster of ions through a collision with another electron. The electron trajectory is shown as a solid blue line, while ions are represented as black dots. The other electrons in the region are not shown.}
    \label{fig:trajectory}
\end{figure}

We also examined individual electron trajectories through space. Such an example electron trajectory is shown in Fig. \ref{fig:trajectory}. 

Because of the random initialization of particle positions, the oscillation damping rate and electron deflection distributions varied from simulation run to simulation run. We averaged over these variations through conducting multiple runs, usually having 6-10 runs per condition of interest with a run time of a few hours for each simulation.

%In our analysis, we did not generally differentiate between free electrons and bound electrons, and included all of the electrons in center-of-mass and deflection angle distribution calculations.  This was in recognition of the fact that Rydberg atoms are continually formed and ionized, that determining a criterion that separates weakly bound and free electrons is problematic, and that the contribution to electron center-of-mass velocity from deeply bound electrons quickly averages to zero, resulting an effectively instantaneous drop in initial velocity that doesn't distort the damping rate extracted from the sinusoidal fits that we used.  We confirmed this to be the case by examining the oscillation damping rate as a function of excluding electrons exhibiting bound state characteristics and seeing no significant change in extracted relative damping rates even while excluding up to 8\% of the electrons that exhibited the strongest bound state character. 

\section{Many-body collisions and electron oscillation damping}

For the first part of our analysis, we used a binary collision operator in the MC simulation that was derived from Rutherford binary Coulomb scattering.\cite{stanton2016ionic} This type of collision operator involves a commonly-used approximate treatment of Coulomb collisions. It has some conceptual and fundamental difficulties, and those limitations will be discussed in the latter part of this article.

For this collision operator, a maximum impact parameter, $b_{\mathrm{max}}$ is set for all electron-ion collisions, regardless of the electron velocity. The total cross section for a collision becomes $\pi b_{\mathrm{max}}^2$. The deflection angle is an analytic function of the impact parameter $b<b_{\mathrm{max}}$ and the electron velocity. In our MC code, if a collision occurs then an impact parameter between 0 and $b_{\mathrm{max}}$ is randomly selected. The electron deflection angle $\chi$ is then equal to $2\cot(e^2/4\pi \epsilon_0 m v^2 b)$, where $v$ is the electron speed. 

We set the value of $b_{\mathrm{max}}=C\lambda_D$, where $\lambda_D=\sqrt{\epsilon_0 k_\mathrm{B} T_e/n e^2}$ is the electron Debye screening length and $C$ is a constant. The value of $C$ was chosen to match that which would be derived from the electron-ion thermalization rate reported in Ref. \onlinecite{dimonte2008molecular}, which was consistent with the effective potential treatment in Ref. \onlinecite{baalrud2012transport}. This value of $C$ was also consistent with classical stopping power predictions in Ref. \onlinecite{grabowski2013molecular}. Note that all of these references involve calculations that include strong coupling effects. In the limit of weak coupling, $C$ was set to be 0.765.

\begin{figure}
    \centering
    \includegraphics[width=0.48\textwidth]{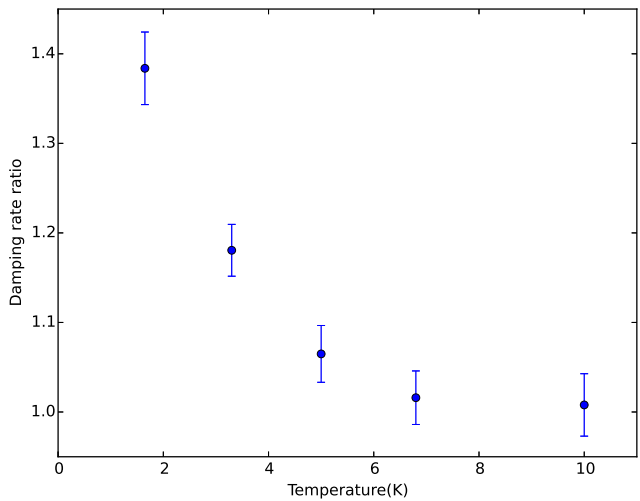}
    \caption{Ratio of full MD-derived oscillation damping rate to the MC-derived oscillation damping rate for the truncated Rutherford collision operator described in the main text as a function of electron temperature. As the electron temperature increases (i.e. for weaker coupling), the agreement between the two different simulation techniques improves. For reference, $\Gamma=0.1$ corresponds to $T_e=5.80\,\mathrm{K}$.}
    \label{fig:ratio_v_Te}
\end{figure}

The ratio between the full MD and MC oscillation damping rates tends toward unity as the electron temperature is increased (i.e. as $\Gamma$ is reduced). Fig. \ref{fig:ratio_v_Te} shows this ratio as a function of $T_e$ and thus electron $\Gamma$. This indicates that while the truncated Rutherford scattering collision operator does not do well at higher values of $\Gamma$, it works well when the electrons are sufficiently weakly coupled. It would of course be possible for the ratio to be made to be equal to one at lower electron temperatures through adjusting $b_{\mathrm{max}}$. Doing so, however, would produce collision operators that did not match the results of Refs. \onlinecite{dimonte2008molecular,baalrud2012transport,grabowski2013molecular} and so would be in contradiction with those works.

The comparison between full MD and MC deflection angle distributions shown in Figs. \ref{fig:histrogram}(a) and \ref{fig:histrogram}(b) was obtained for conditions where $\Gamma$ is higher ($\Gamma$=0.35). In these figures, it is evident that the MC simulation results in too high a ratio of small angle to large angle electron deflections as compared to the full MD. In principle, this could simply be due to an inaccurate collision operator. Through altering the value of $b_{\mathrm{max}}$ and artificially altering the value of $e$ in the collision operator calculation, it was possible to get good agreement between the angle distributions in both the MC and full MD simulations (i.e. to match curves of the type shown in Fig. \ref{fig:histrogram}(b)). Once this was accomplished, however, the MC-derived damping rate became larger than the full MD-derived one by a factor of 3, meaning that matching the deflection angle distributions in this way produced a radical increase in the disagreement in the oscillation damping rate obtained from the two simulations. We also altered the $b_{\mathrm{max}}$ parameter alone to adjust the MC oscillation damping to match the full MD case. In that instance, the angle distribution curves still did not match, again indicating the insufficiency of such alterations of the binary collision operator to simultaneously match the angle distributions and oscillation damping rates.

In principle, such a disagreement between oscillation damping rate and deflection angle distribution could occur if there were sufficiently deeply-bound Rydberg atoms that were present before the oscillation was initiated and remained throughout the time that the oscillation damped. We confirmed that was not the cause of the discrepancy by performing auxiliary simulations where deeply-bound Rydberg atoms' ions and electrons were removed. Removing these deeply-bound Rydberg atoms did not substantially impact the oscillation damping rate.  Additionally, such Rydberg atoms would be expected to decrease the amplitude but not the decay time constant of the oscillation curves, further indicating that deeply bound Rydberg atoms were not the source of the observed discrepancies.

To investigate the role of MBTBS contributions to the oscillation damping rate, we performed a series of simulations where we altered the ratio between the binary electron-ion collision rate and the MBTBS rate. This was accomplished by artificially decreasing the ion charge while increasing the ion number to keep the charge density of the ions constant. The ratio of binary electron-ion collision rate to MBTBS rate changes because the MBTBS rate scales more steeply with the ion charge. This can be inferred from expressions for the three-body recombination rate\cite{goforth1976recombination} as well as noting that many-body collisions involve volume considerations while binary collisions scale as a cross-sectional area, and so changes in fundamental length scale via altering the ion charge impact many-body more than binary collisions. This was confirmed in the full MD simulations through the observation of a reduced TBR heating rate for smaller ion charges. This, by the way, is why we did not conduct this test with higher ion charge values. Increasing the charge led to more TBR heating such that the desired values of electron $\Gamma$ could not be obtained.

\begin{figure}
    \centering
    \includegraphics[width=0.48\textwidth]{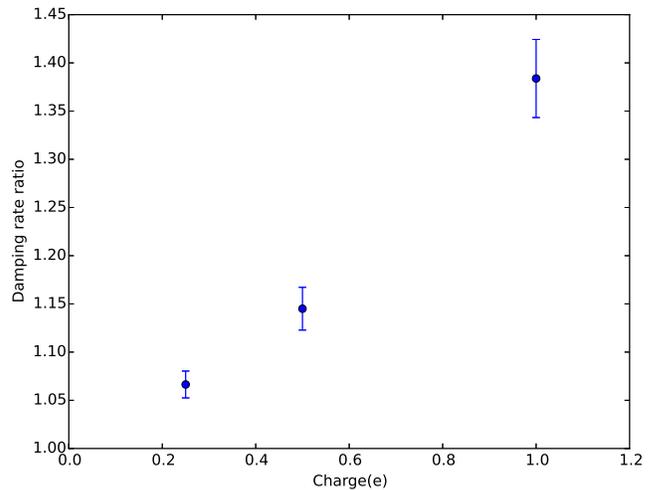}
    \caption{Ratio of oscillation damping rates determined from the full MD simulation to the MC simulation as a function of individual ion charge. For all simulations, as the charge of the individual ions were altered, the ion number density was adjusted to keep the charge density fixed. As the ion charge tends toward zero, the ratio between the full MD and MC oscillation damping rates approaches unity.
    %The collision operator used for the MC simulation was the truncated Rutherford scattering operator discussed in the main text.
    }
    \label{fig:scaled_charge}
\end{figure}

The results of these simulations are shown in Fig. \ref{fig:scaled_charge}. Note that as the ion charge is decreased, the ion number has to increase and the simulations become more computationally expensive. Similar to the case of weak-coupling, as the individual ion charge is decreased the discrepancy between the full MD and MC results diminishes. This suggests that as the ratio of many-body to binary collisions decreases, the disagreement between the full MD and MC results decreases as well. We note that in these simulations nothing changes about the electron $\Gamma$ or electron density --- only the ion parameters change.

We performed an additional analysis to characterize the number of electrons in bound states in the simulations. We defined a ``localization'' parameter for each electron over a chosen time period $\Delta t$. The localization parameter for an individual electron, which is a purely classical quantity, is defined as the ratio of the net spatial displacement of the electron $\Delta S$ to the average velocity of the electron in the time period, $\bar{v}$, times $\Delta t$. In the absence of any collisions, this ratio would be one. Electron-electron collisions cause velocity changes and deflections and so those collisions produce a distribution of localization parameters. Binary electron-ion collisions cause further deflections. Bound state electrons, however, have much smaller displacements than free electrons in general and so for those electrons their localization parameters are much smaller.

\begin{figure}
    \centering
    \includegraphics[width=0.48\textwidth]{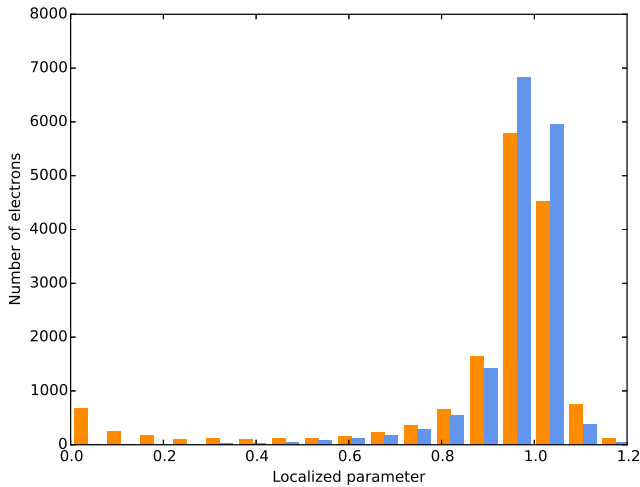}
    \caption{Histogram plot of the localization parameter. Blue bars represent the MC simulation results and the orange bars represent MD results. The MD results show a much higher fraction of electrons with localization parameters indicating a high degree of localization (i.e.\ small values as defined in the main text). The time period used for the data in this plot was 5 $\mathrm{ns}$. }
    \label{fig:localize}
\end{figure}

Fig. \ref{fig:localize} shows the distribution of localization parameters for electrons in the full MD and MC simulations. There are a far larger number of highly localized electrons in the full MD case. A few of these represent relatively deeply bound Rydberg atoms, but many more represent electrons that spend some time in a weakly bound state where they acquire large velocity angle deflections in a relatively short period of time. 5\% of the electrons have a localization parameter of about 0.1 or less, indicating that a substantial fraction of the electrons are significantly localized in the plasma. At this degree of localization, an electron's velocity direction would be nearly completely randomized in a plasma oscillation period. The localization parameter distribution and the observed gap in the oscillation damping rate between the full MD and MC simulations are thus in general agreement with one another.

\section{Discussion}

The fraction of electrons that have small localization parameters combined with the fact that electrons can be scattered into bound states and then scattered out of them in timescales on the order of a few plasma oscillation periods\cite{bannasch2011rydberg} suggests that MBTBS collisions are largely responsible for the additional observed damping rate in the full MD simulation as compared to the MC simulations at higher electron $\Gamma$.

There are other possibilities, however, that were considered. We examined finite ion mass effects, non-spherical symmetry, and finite spatial extent effects and found them to have no more than a few percent effect on the oscillation damping rate. Ion correlations were found to have up to a several percent impact on the oscillation damping rate, presumably through ion screening effects. However, for the infinite mass and randomly placed ions in the current set of simulations such considerations are not relevant.

Perhaps the most important consideration, however, concerns the use of the truncated Rutherford collision operator. Such a collision operator is easy to treat analytically and easy to implement. It is, however, unphysical in two respects. First, there is no screening included in the collision calculation other than the truncation. Second, the truncation of the impact parameter is quite unphysical in that it is not that a finite-ranged potential interaction is used, but rather a potential that is finite in radial extent prior to the collision and infinite after it.\cite{stanton2016ionic} It is true that the truncated Rutherford collision operator matched the full MD simulation well under conditions of weak coupling and lower MBTBS-to-binary-collision ratios, but there are reasons to believe that such agreement would likely break down when strong coupling physics became more significant. In addition, recent work indicates that there is a Barkas-like effect in collisions between unlike-signed particles (such as electrons and positively-charged ions)\cite{shaffer2019barkas} that becomes significant with increasing $\Gamma$, and such an effect is not present in the Rutherford scattering operator.

Following the work in Ref. \onlinecite{shaffer2019barkas}, we used a collision operator that included screening effects and the Barkas-like effect as derived from an effective potential to compute the predicted oscillating damping rate at the $\Gamma$=0.35 conditions. We obtained the collision differential cross sections necessary for the collision operator from the authors of Ref. \onlinecite{shaffer2019barkas}. Using this collision operator in the MC simulation, we got some improvement as the MC-derived oscillation damping rate increased by 6\% over the truncated Rutherford collision operator. This increase is not enough to explain the original 37\% full MD discrepancy, however. We note that using a simple Yukawa interaction potential with a screening length chosen to match the stopping power and thermalization rates in Refs. \onlinecite{dimonte2008molecular,baalrud2012transport,grabowski2013molecular} actually reduced the MC predicted damping rate by 20\% as compared to the truncated Rutherford calculation. So the treatment that included the Barkas effect and used an effective potential was much better than a simple implementation of a Yukawa potential. In the end, though, improvements in the MC collision operator did not resolve the full discrepancy with the full MD calculation for the $\Gamma=0.35$ condition.

In light of this, to further investigate whether rare events that result in a very high deflection of an electron's velocity direction are the cause of the discrepancy, we conducted angle distribution calculations using the usual MC collision operator plus an addition. We added the possibility of another -- artificial -- collision occurring where there was a probability in a time period for each electron whose speed was below a maximum speed ($v_{\mathrm{max}}$) to have its velocity direction completely randomized. This additional artificial type of collision was chosen to roughly simulate an electron becoming bound, deflecting in direction substantially while bound, and then becoming unbound via a collision after that deflection. The use of a threshold was motivated by the fact that MBTBS collisions are more likely for slower-moving electrons.  We adjusted the random probability for these artificial collisions until the oscillation damping decay rate in the MC simulation matched that in the MD simulation.

\begin{figure}
    \centering
    \includegraphics[width=0.48\textwidth]{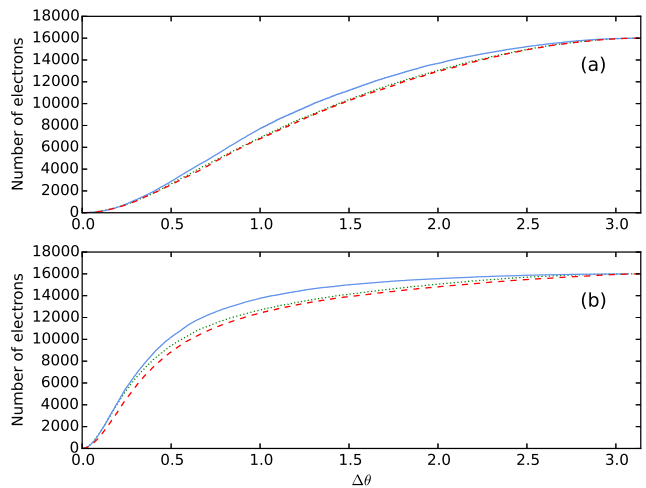}
    \caption{Cumulative collision angle distribution plots, similar to those in Fig. \ref{fig:histrogram}(b). The simulation parameters such as density and temperature are the same in this plot as in that one. Part (a) in this plot represents results obtained with $\Delta t=20\,\mathrm{ns}$, while part (b) corresponds to $\Delta t=5\,\mathrm{ns}$. The blue solid line represents the usual MC result as presented in the text, while the red dashed line represents the MD result. The green dotted line represents the MC result with the addition of 
    %with randomized velocity rate at 11 electron per time step but only randomized the electron have a velocity smaller than 8000m/s within this 11 electrons
    randomly occurring events where the velocity direction of an electron is completely randomized.%See the main text for details.
    }
    \label{fig:matching}
\end{figure}

The resulting angle distribution curves are shown in Fig. \ref{fig:matching}. The random probability rate for the artificial high-angle-deflection collisions was set to be $2.42\times10^5\,\mathrm{s}^{-1}$ with $v_{\mathrm{max}}=8\,\mathrm{km/s}$. It is evident that in the upper plot in Fig. \ref{fig:matching} the additional artificial collision probability being added to the MC simulation reproduces the angle distribution curve well for $\Delta t$=20 ns. The comparison between the modified MC and MD is not perfect, as seen in the lower plot of Fig. \ref{fig:matching} for a shorter $\Delta t$. However, this comparison shows that the MD simulation can be matched over $\Delta t=20 \mathrm{ns}$ in both oscillation damping rate and angle deflection distribution by adding rare but very high-deflection events that would be similar to the net expected effect from actual MBTBS collisions.

It is hard to definitively eliminate the possibility that yet another binary collision operator could reproduce the rates observed in other work\cite{dimonte2008molecular,baalrud2012transport,grabowski2013molecular}, could match the angle distributions between full MD and MC simulations while simultaneously matching the oscillation damping rates, could produce the same degree of electron localization, would reduce to the truncated Rutherford collision operator in the limit of small individual ion charge, and would be physically reasonable. The balance of all of these considerations and results presented above are highly consistent with the majority of the gap between the full MD and MC predictions being due to MBTBS collisions and subsequent scattering of those electrons back into unbound states.

\section{Conclusion}

We have used two separate simulation techniques to calculate the damping rate of electron oscillations in conditions where electron strong coupling is relevant. We found that the result obtained from a full MD simulation that treated both the electrons and ions as point particles did not agree at the tens of percent level with a MC simulation that treated electron-ion collisions via a random binary collision operator based on an electron-ion collision theory consistent with work published in Refs. \onlinecite{dimonte2008molecular,baalrud2012transport,grabowski2013molecular}. We ascribe this difference to the effect of many-body collisions that scatter electrons into being bound to one or more ions. When these electrons become bound, their velocity direction changes quickly. These electrons can be scattered back into a unbound state and thus can contribute significantly to the decay rate of the average electron momentum in the plasma. Scaling studies and characterizations of the degree of ``localization'' of the electrons indicated that these many-body collisions to bound states are sufficient to observe the explained simulation discrepancy.

In most plasmas, three-body recombination is not the dominant recombination mechanism and can generally be ignored. This won't be the case for sufficiently strongly coupled electrons, however, and MBTBS collisions like those discussed here will be relevant. This is the case, for instance, for some warm dense matter conditions.\cite{cho2016measurement,dornheim2018uniform} Also, recombination effects that impact electron-ion collision properties are relevant to plasma stopping power.\cite{deutsch2018ion} We note that significant discrepancies have manifested themselves at a relatively mild value of $\Gamma$=0.35, indicating that even with only a moderate degree of strong coupling, many-body collisions need careful consideration in order to avoid distorting binary collision predictions or analyses.  

\section{Acknowledgements}

This work was supported by the Air Force Office of Scientific Research (AFOSR), Grant No. FA9550-17-1-0148. We also acknowledge Craig Witte's efforts in the initial creation of the molecular dynamics simulation code used in this work. In addition, we acknowledge useful conversations with Nathaniel Shaffer and Scott Baalrud and their sending us detailed collision predictions based on their recent work.\cite{shaffer2019barkas}

\bibliography{UCP.bib}
\end{document}